\documentclass[A4paper,12pt]{article}

\usepackage{fancyhdr}
\usepackage{anysize}

\marginsize{1.5cm}{1.5cm}{0cm}{0.5cm}

\usepackage{graphicx}
\usepackage{amsmath}
\usepackage{amsfonts}
\usepackage{epsfig}
\usepackage{color}
\usepackage[utf8]{inputenc}
\usepackage{hyperref}

\date{}
\begin{document}

\title{ Asymptotic Boundary Conditions and Square Integrability in the Partition Function of AdS Gravity}
\author{Joel Acosta$^{\dag}$, Alan Garbarz$^{\ddag}$, Andr\'es Goya$^{\ddag}$, Mauricio Leston$^{\mathsection}$}
\maketitle
\vspace{.25cm}

\begin{minipage}{.9\textwidth}\small \it 
	\begin{center}
    $^\dag$ Departamento de  Matem\'atica-FCEN-UBA \& IMAS-CONICET,
	Ciudad Universitaria, pabell\'on 1, 1428, Buenos Aires, Argentina.
     \end{center}
\end{minipage}

\vspace{.25cm}

\begin{minipage}{.9\textwidth}\small \it 
	\begin{center}
    $^\ddag$ Departamento de F\'isica-FCEN-UBA \& IFIBA-CONICET,
	Ciudad Universitaria, pabell\'on 1, 1428, Buenos Aires, Argentina.
     \end{center}
\end{minipage}

\vspace{.25cm}

\begin{minipage}{.9\textwidth}\small \it \begin{center}
    $^\mathsection$ Instituto de Astronom\'ia y F\'isica del Espacio (IAFE),
	Pabell\'on IAFE-CONICET, Ciudad Universitaria, C.C. 67 Suc. 28, Buenos Aires, Argentina.
     \end{center}
\end{minipage}

\vspace{.5cm}

\begin{abstract}
\small
There has been renewed interest in the path-integral computation of the partition function of AdS$_3$ gravity,  both in the metric and Chern-Simons formulations. The one-loop partition function around Euclidean AdS$_3$ turns out to be given by the vacuum character of Virasoro group. This stems from the work of Brown and Henneaux (BH) who showed that, in AdS$_3$ gravity with sensible asymptotic boundary conditions, an infinite group of (improper) diffeomorphisms arises which acts canonically on phase space  as two independent Virasoro symmetries. The gauge group turns out to be composed of so-called ``proper'' diffeomorphisms which approach the identity at infinity fast enough. However, it is sometimes far from evident to identify where BH boundary conditions enter in the path integral, and much more difficult to see how the improper diffeomorphisms are left out of the gauge group. In particular, in the metric formulation,  Giombi, Maloney and Yin obtained the one-loop partition function around thermal AdS$_3$ resorting to the heat kernel method to compute the determinants coming from the path integral.  Here we identify how BH boundary conditions follow naturally from the usual requirement of square-integrability of the metric perturbations. Also, and equally relevant, we clarify how the quotient by \emph{only} proper diffeomorphisms is implemented, promoting the improper diffeomorphisms to symmetries in the path integral. Our strategy is general enough to apply to other approaches where square integrability is assumed. Finally, we show that square integrability implies that the asymptotic symmetries in higher dimensional AdS gravity are just isometries.  

\vspace{.5cm}

\begin{flushleft}
\hrulefill\\
\footnotesize
{E-mails: joel@dm.uba.ar, alan@df.uba.ar, af.goya@df.uba.ar, mauricio@iafe.uba.ar}
\end{flushleft}

\end{abstract}

\newpage

\section{Introduction}

Three dimensional pure gravity with negative cosmological constant has been widely used as a toy model for addressing problems of quantum gravity (see for example \cite{Witten:3D.GR.Chern.Simons, Witten:3D.Grav.Revisited, Maloney.Witten,  Kim.Porrati:1508}).  One important step towards the understanding of the quantum description of AdS$_3$ gravity was taken in \cite{Maloney.Witten}, where it was computed the  partition function,
\begin{equation}
	Z(\beta,\theta)=\text{Tr}\left(e^{-\beta H-i \theta J} \right) \,,
\end{equation}
with $H$ and $J$ the Hamiltonian and  angular momentum operators respectively, as a functional integral over Euclidean 3 dimensional geometries with conformal boundary a two-torus with modular parameter $\tau=\theta+i\beta$. The symmetry group of the classical space of solutions played a prominent role in that work, as we shall explain.

In 3 dimensional gravity  there are no local degrees of freedom, and then it could be expected there is a trivial phase space formed just by AdS$_3$ (for negative cosmological constant, which we assume henceforth). However it is known that there are interesting solutions like black holes \cite{BTZ,BTZ:Geometry}, which are locally but not globally AdS$_3$, and accordingly in \cite{Maloney.Witten} the partition function was decomposed as a sum over different saddle points. Brown and Henneaux \cite{Brown.Henneaux:AdS3} showed that the asymptotic boundary conditions and related asymptotic symmetry group are responsible  for a rich phase space of solutions, even in the sector of small perturbations around AdS$_3$.  The asymptotic symmetry group is given by two copies of Diff$(S^1)$, that however are centrally extended when realized through the algebra of charges, giving two copies of the Virasoro group with equal central charges $c=3\ell/2G$ \cite{Brown.Henneaux:AdS3}. \cite{Maloney.Witten} relied heavily on this result. Such enhancement of symmetries does not occur in higher dimensions where the asymptotic symmetry group coincides with the isometry group\footnote{This can be understood from the fact that the asymptotic symmetry group is the same as the conformal symmetry group of the boundary metric. For $D>3$ it is the finite dimensional group $SO(D-1,2)$ in Lorentzian signature or $SO(D,1)$ in Euclidean signature. The $D=4$ case can be explicitly seen in \cite{Henneaux:AdS4}.}. The asymptotic symmetries of AdS are nowadays understood, in the context of AdS/CFT \cite{Maldacena:AdS.CFT}, as the local conformal group of the boundary theory. 

The setting of \cite{Brown.Henneaux:AdS3} is such that i) the asymptotic  boundary conditions on the metric ensure having conserved charges that are finite, and ii) any diffeomorphism that approaches the identity at infinity fast enough not to change any conserved charge, is considered a redundancy and must be gauged away; these are the so-called ``proper'' diffeomorphisms (diff). The remaining diffeomorphisms, modulo proper diffeomorphisms, are called ``improper'' and modify conserved charges. These improper diffs generate the so-called boundary gravitons and are  symmetries of phase space acting through the coadjoint action of the Virasoro group (see \cite{ Garbarz.Leston:1403,Barnich:2014zoa} and references therein).

 Taking into account the aforementioned structure of the classical phase space around (global) AdS$_3$ space, it was argued in \cite{Maloney.Witten} that the Hilbert space of the quantum theory is bigger than just the vacuum state $\Omega$. The corresponding Hilbert space $\mathcal{H}_0$ is the Verma module of the vacuum, i.e. the irreducible representation of the Virasoro algebra constructed by acting with the Virasoro generators $L_{-n}$ on  $\Omega$. This can be thought as acting with Virasoro generators on the state associated with the AdS$_3$ solution. Taking into account that $L_0+\bar{L}_0$ is the energy and $L_0-\bar{L_0}$ the angular momentum, the torus partition function coincides then with the vacuum character of two copies of the Virasoro group Tr${}_{\mathcal{H}_0}(q^{L_0}\bar{q}^{\bar{L}_0})$:
\begin{equation}\label{character}
Z(\beta,\theta)|_{{\rm AdS}_3\,{\rm sector}}=|q|^{c/12} \prod_{n>1} \frac{1}{|1-q^n|^2} \,,
\end{equation}  
where $q=e^{2\pi i \tau}$ with $\tau=\theta+i \beta$ (both parameters are real and represent the angular potential and the inverse temperature). Expanding the character as a sum of powers of $q$ and $\bar{q}$, one can identify the contribution of the vacuum as $|q|^{c/12}$, and all the other terms are contributions of the Virasoro descendants. The effective action $S_{\rm eff}=-{\rm log}(Z)$ is of the form  $c(S_0 + \frac{1}{c} S_1)$, where $S_0$ is just the Einstein-Hilbert action evaluated in Euclidean AdS$_3$. Since $1/c$ plays the role of $\hbar$ in QFT, the previous result shows that the effective action does not have higher order contributions; for that reason, a QFT direct computation must give a 1-loop exact result. 

The one-loop partition function can be formulated in terms of an Euclidean path integral where the action is perturbed around AdS$_3$ up to second order (we will expand below). The final expression is then expressed as a quotient of determinants of some operators. This holds both in the metric as well as in the Chern-Simons formulations of AdS$_3$ gravity \cite{Witten:3D.GR.Chern.Simons,Achucarro.Townsend:3D.Sugra.Chern.Simons}. In the latter case, the partition function has been extensively studied recently in \cite{Porrati:2019knx}. Here we are interested in the metric formulation of gravity where the determinants correspond to Laplace operators on fields of integer spin $0$, $1$ and $2$. They have been computed to obtain the 1-loop partition function,  by means of a heat kernel procedure in  \cite{Giombi:One.Loop} (see also \cite{David:2009xg}), and by using a quasi-normal mode method in \cite{Castro:2017mfj} which can be related to scattering poles through  the Selberg zeta function (see \cite{Martin:2019flv} and references therein). 

Let us be more precise. Consider the action of General Relativity with negative cosmological constant and expand around AdS$_3$ as $g=g_{\rm AdS}+h$, keeping terms up to second order in the perturbation $h$. Since local diffeomorphisms are gauge symmetries, a gauge-fixing is required together with auxiliary ghost fields. The metric perturbation $h$ can be described by two fields on AdS$_3$, a scalar field (its trace) and a traceless symmetric rank-two field, while the ghosts are vector fields. Since the action is Gaussian in the fields, the result is a ratio of determinants, one for each field. In \cite{Giombi:One.Loop} these are computed by the heat kernel method, with a prescribed regularization. The thermal feature of the geometries\footnote{This means that Euclidean time is identified with a translation given by $2\pi\beta$ while the angular coordinate is rotated by $2\pi\theta$. An elegant way of seeing such geometry is by considering the quotient $\mathbb{H}^3/\Gamma$, where $\Gamma$ is the Fuchsian group generated by diag$(q^{1/2},q^{-1/2})\in$ SL$(2,\mathbb{C})$.} is taken into account in the path integral by means of the method of images, therefore obtaining the heat kernels of thermal AdS$_3$.
Having these, the last step is to gather all together and the Virasoro character (\ref{character}) is obtained.  

A natural question that arises is: where are implemented  the following two important ingredients of the classical phase space in the  heat kernel computation?
\begin{itemize}
\item The imposition of asymptotic boundary conditions of Brown-Henneaux.

\item The promotion of improper diffeomorphisms to symmetries as opposed to gauge redundancies.

\end{itemize}
This is the question that we will answer. Let us recall that the last point was responsible for the rich structure of the Hilbert space leading to the expression (\ref{character}) in terms of Virasoro characters.

Recently, in \cite{Witten:Lect.Bdy.Conditions} it was remarked that the square integrability of the fields together with appropriate boundary conditions are crucial to describe Euclidean gravity as an elliptic boundary value problem \cite{Anderson_2008}. In the present context of AdS$_3$ gravity,  we will see in the next sections that the square integrability of the fields, assumed in the computations of the heat kernels,  will lead both to the (Euclidean) asymptotic conditions of Brown-Henneaux and will automatically exclude the  improper diffeomorphisms from the quotient by gauge redundancies, for which the ghost field are introduced. In this way, the square integrability condition turns out to be the root of the emergence of Virasoro group as symmetries in the QFT computation in \cite{Giombi:One.Loop}.  
An important observation is that our presentation is actually not restricted to the heat kernel method, but actually holds for any method used to calculate determinants within a space of square-integrable fields. 

Finally, we will apply the same reasoning for Euclidean AdS$_D$ gravity, with $D>3$, and we will obtain that the asymptotic symmetry group is nothing but the group of isometries, in agreement with \cite{Henneaux:AdS4}.

\section{Metric perturbations}\label{metric3d}

We are interested in understanding the phase space of metric perturbations  that is relevant for the computation of the one-loop partition function of AdS$_3$ gravity. 
The partition function of gravity is roughly of the form $\int Dg \, e^{-\int d^3x \sqrt{g} (R+2)  }$. The integrand in the exponent should be expanded around the hyperbolic metric  up to second order and this gives an action of the form $\int h L h$ with $L$ a non-elliptic second order differential operator (see \cite{Witten:Lect.Bdy.Conditions} for more details). We henceforth call $h$ to the metric perturbation around AdS, so we have $\delta g=h$ and from now on $g:=g_{\rm AdS}$ is the hyperbolic metric. We also use the notation $\mathbb{H}^3$ for Euclidean AdS$_3$ or hyperbolic space and set the AdS radius to one. The functional integral should sum over all those $h$ that satisfy a sensible boundary condition and belong to some judicious gauge choice. We will scrutinize these requirements in this and the following Section.

A natural space where the operator $L$ acts is the space of square integrable rank 2 symmetric tensors. So we consider that this is the space to which $h$ belongs.  The inner product on metric perturbations is
\begin{equation}
     \langle  h \mid h' \rangle = \int_{\mathbb{H}^3} d^3x \sqrt{g}\, \bar{h}^{\mu \nu} h'_{\mu \nu} = \int_{\mathbb{H}^3} d^3x \sqrt{g}\,  g^{\mu \alpha} g^{\nu \beta}\bar{h}_{\mu \nu} h'_{\alpha \beta} \,.
     \label{eq:L2_metricas}
\end{equation}
If we use Poincar\'e coordinates where $ds^2=\frac{dx^2+dy^2+dz^2}{z^2}$, the norm of $h$ is,
\begin{equation}
    \langle h | h \rangle = \int_{\mathbb{H}^3} dx\,dy\,dz\, z \left[ h^2_{xx} + h^2_{yy} + h^2_{zz} +2h^2_{xy} + 2h^2_{xz}+ 2h^2_{yz} \right] \,,
    \label{eq:producto_interno_h}
\end{equation}
from where we conclude\footnote{We are neglecting analyzing the dependence of $h$ on the boundary coordinates, since we are implicitly assuming the boundary is the Riemann sphere or a quotient of it.} that $ h_{\mu\nu} \sim z^{\epsilon-1}, \epsilon > 0,$ for $z\rightarrow 0 $. This fall-off behavior coincides with the boundary conditions of Brown and Henneaux, with the exception of the components $h_{x z}$ and $h_{y z}$ which here are of order $\mathcal{O}(1)$ versus $\mathcal{O}(z)$ of \cite{Brown.Henneaux:AdS3}. We will show that these components are actually pure gauge. 

Some of the perturbations come from diffeomorphisms, namely $h=\mathcal{L}_\xi g$, with $\xi$ some vector field. Naively one is tempted to mod out by all of these perturbations, however that would not be correct, as we shall see. Let us call $\mathcal{V}$ the space of vectors that generate square-integrable perturbations, namely
\begin{equation}
	\mathcal{V}=\lbrace \xi \in {\rm Vect}(\mathbb{H}^3) \mid \mathcal{L}_{\xi}g \in L^2 \rbrace \,.
	\label{eq:def_V}
\end{equation}
Then, it can be shown that the most general vector field in $\mathcal{V}$ can be written
\begin{equation}
    \begin{aligned}
    \xi^x &= U(x,y) -\frac{z^2}{2} \partial^2_{xx} U(x,y) + \mathcal{O}(z^3) \,, \\
    \xi^y &= V(x,y) - \frac{z^2}{2} \partial^2_{yy} V(x,y) + \mathcal{O}(z^3) \,, \\
    \xi^z &= \dfrac{z}{2} \left(\partial_x U(x,y)+\partial_y V(x,y) \right) +\mathcal{O}(z^3) \,,
    \end{aligned}
    \label{eq:forma_vectores}
\end{equation}
with the Cauchy-Riemann conditions $\partial_x U = \partial_y V$, $ \partial_y U = - \partial_x V$. Of course, these equations imply that we can form holomorphic and anti-holomirphic functions. Defining the complex coordinate $w=x+iy$ the vectors read

\begin{equation}
    \begin{aligned}
    \xi^w &= f(w) -\frac{z^2}{2} \bar{\partial}^2 \bar{f} + \mathcal{O}(z^3) \,, \\
    \xi^{\bar{w}} &= \bar{f}(\bar{w}) - \frac{z^2}{2} \partial^2 f + \mathcal{O}(z^3) \,, \\
    \xi^z &= \dfrac{z}{2} \left(\partial f +\bar{\partial} \bar{f} \right) + \mathcal{O}(z^3) \,.
    \end{aligned}
    \label{vectores_bh_complejas}
\end{equation}
These are the usual Brown-Henneaux asymptotic vector fields, with the exception of a milder fall-off condition $\mathcal{O}(z^3)$ instead of $\mathcal{O}(z^4)$ in the non-radial components. The vectors with $f=0$ and $\bar{f}=0$ are called ``proper'' diffeomorphims and generate a pure gauge deformation of the background metric, since they do not modify the conserved charges \cite{Brown.Henneaux:AdS3} (we will further discuss this point in the next Section). The so-called ``improper'' diffeomorphisms are the vectors \eqref{vectores_bh_complejas} modulo the proper ones. The components $h_{w z}$ and $h_{\bar{w} z}$ are only affected by the proper vectors, therefore can be considered as pure gauge as we mentioned before. The asymptotic vector fields \eqref{vectores_bh_complejas} are in agreement with those of \cite{Barnich.Brandt}, and this guarantees both finite conserved charges and finite central charge (in the Lorentzian case). 

We should emphasize that in order to obtain the Virasoro symmetry what it really matters is how the algebra of vectors is represented on the space of metrics, and not the precise form of the vectors or the metric components. The algebra formed by these vectors is best displayed by passing to a mode decomposition. We define $L_m$ as the vector with $(f=-w^{m+1},\bar{f}=0)$, and $\bar{L}_m$ the vector with $(f=0,\bar{f}=-\bar{w}^{m+1})$. Then we have two copies of the Witt algebra,
\begin{equation}
	\begin{aligned}
		\left[L_m,L_n\right] &= (m-n)L_{m+n} \,, \\
		\left[\bar{L}_m,L_n\right] &= 0 \,, \\
		\left[\bar{L}_m,\bar{L}_n\right] &= (m-n)\bar{L}_{m+n} \,.
	\end{aligned}
\end{equation}
As is well-known, the way these vector fields act on the hyperbolic metric is by the shift
\begin{equation}
	h_{ww}=0\mapsto h_{ww}=-\frac{1}{2}\partial^3 f \,,
\end{equation}
and analogously for its complex conjugate. This indicates that the vectors act with the coadjoint representation of the Virasoro algebra.

\section{Ghosts and asymptotic vector fields}

We shall now analyze in more detail vectors (\ref{vectores_bh_complejas}). The reader should keep in mind that ghost fields generate gauge symmetries and will eventually  be introduced with their corresponding action in the path integral, so they need to be square-integrable too. The first thing to note is that only the proper diffs are square integrable. This follows from the inner product on vector fields, 
\begin{equation}
    \langle \xi| \xi' \rangle = \int_{\mathbb{H}^3} d^3x \sqrt{g}\, \bar{\xi}^\mu \xi'_{\mu} = \int_{\mathbb{H}^3}d^3x \sqrt{g}\, \bar{\xi}^\mu \xi'^\nu g_{\mu \nu} \,.
    \label{eq:L2_vectores}
\end{equation}
On the contrary, the leading and sub-leading terms in (\ref{vectores_bh_complejas}) are not square integrable. Thus, we arrive to an important result: the ghosts in the path integral are precisely the square-integrable part of (\ref{vectores_bh_complejas}) which coincide with proper diffeomorphisms. This implies that perturbations of the metric $\mathcal{L}_{\xi}g$, with $\xi$ generating an improper diff, will not be gauged away. This leaves out of the gauge redundancies the improper diffs, and then they survive as symmetries and contribute to the partition function. 

So far we have not mentioned a specific gauge choice. However, it is crucial to understand the role that a gauge choice plays in this analysis: given that an improper diff does not generate a gauge redundancy it should not break the gauge choice. %
On the contrary, when $\xi$ is a generator of a proper diff then it should break the gauge. Let us call $\mathcal{V_P}$ and $\mathcal{V_I}$ the subspaces of generators of proper and improper diffeomorphisms. Notice that\footnote{They are expressed in a coordinate system that leaves out the region $z=\infty$, should this point be added to perform a compactification.} $\mathcal{V_I}=\mathcal{V}/\mathcal{V_P}$.

In order to study the problem just raised, let us use the de Donder gauge fixing condition: $T(h):=\nabla^\nu h_{\mu\nu}-{\frac{1}{2}} \partial_\mu \text{Tr}h=0$. This is a good gauge-fixing condition, since to any perturbation $h_{\mu\nu}$ with $T(h)\neq 0$ we can add $\mathcal{L}_\xi g$, such that $h'=h+\mathcal{L}_\xi g$ now satisfies $T(h')=0$. This is granted by the existence of a unique solution to $P(\xi)=T(h)$, where $P(\xi)_\mu:=-\nabla^2\xi_\mu-R_{\mu\nu}\xi^\nu$ is an invertible operator \cite{Witten:Lect.Bdy.Conditions}. Note, however, that $P$ is an invertible operator only within a fixed space of vectors, which for simplicity we take to be the square-integrable vectors\footnote{For a more technical study we suggest \cite{Anderson_2008}. }. We will show soon that \emph{there are} non-square-integrable vector fields in the kernel of $P$. 

Now, the usual BRST procedure is to add  a gauge-fixing action and a ghost action (see for example  \cite{Giombi:One.Loop}). The latter, calling $\eta$ the ghost fields, is
\begin{equation}
    S_{\rm ghost}=\dfrac{1}{32 \pi G }\int d^3x\sqrt{g}\,  \bar{\eta}_{\nu}\left(-g^{\mu \nu}\nabla^2 - R^{\mu \nu}\right)\eta_{\nu}=\dfrac{1}{32 \pi G }\int d^3x\sqrt{g}\,  \bar{\eta}^{\nu}P(\eta)_{\nu} \,.
\end{equation}
These $\eta$ vectors implement the gauge transformation $\mathcal{L}_\eta g$, and should be square-integrable since are to be integrated in the action. They are then the elements in $\mathcal{V_P}$. What should be checked is that $P(\mathcal{V_P})\neq 0$, which is straightforward to show (it is also guaranteed by the invertibility of $P$). Physically this means that generators of proper diffs change the gauge condition. On the other hand, it is a fact that the representative $\xi \in \mathcal{V_I}$ which is written as in \eqref{vectores_bh_complejas} with no $\mathcal{O}(z^3)$ satisfies $P(\mathcal{\xi})=0$. Namely, that improper vector fields \emph{do not change the gauge fixing condition}, and thus can be safely considered symmetries. Note that we are not contradicting the claim that $P$ is invertible, since this holds for square-integrable vector fields (or may be a technically broader space), but elements of $\mathcal{V_I}$ are not square integrable.

\section{Square integrability in higher dimensions}

In order to explore in higher dimensions which is the asymptotic symmetry group coming from square integrability, let us repeat what we have done in Section \ref{metric3d}. The inner product is now, 
\begin{equation}
     \langle  h \mid h' \rangle = \int_{\mathbb{H}^D} d^Dx \sqrt{g}\, \bar{h}^{\mu \nu} h'_{\mu \nu}   = \int_{\mathbb{H}^D} d^Dx \sqrt{g}\,  g^{\mu \alpha} g^{\nu \beta}\bar{h}_{\mu \nu} h'_{\alpha \beta} \,.
     \label{eq:L2_metricasD}
\end{equation}
This implies that in Poincaré coordinates, the norm of $h$ is,
\begin{equation}
    \langle h | h \rangle = \int_{\mathbb{H}^D} d^d x\, dz\, z^{3-d} \left[h_{zz}^2+\sum_{i}^d \left( h^2_{ii}  + 2 h^2_{i z}\right)+\sum_{i,j=1}^d h^2_{ij} \right] \,,
    \label{eq:producto_interno_hD}
\end{equation}
where we have defined $d=D-1$. From this expression we conclude that $ h_{\mu\nu} \sim z^{\epsilon-2+d/2}, \epsilon > 0,$ for $z\rightarrow 0 $. Then the most general vector field in the higher-dimensional analog of $\mathcal{V}$ is of the form, 

\begin{equation}
    \begin{aligned}
    \xi^z&=\dfrac{z}{d}\left(\partial_{x^1} U^1(x) + \cdots + \partial_{x^d}U^d(x) \right) +\mathcal{O}\left(z^{\lfloor\frac{d+4}{2}\rfloor}\right) \,, \\
    \xi^{i}&=U^i(x) - \frac{z^2}{2}\dfrac{\partial^2 U^i}{\partial x^i\partial x^i} + \mathcal{O}\left(z^{\lfloor\frac{d+4}{2}\rfloor}\right) \,,
    \end{aligned}
    \label{eq:asym_vectorsD}
\end{equation}
where $\lfloor a \rfloor$ is the integer part. The $U^i$ are functions of the boundary coordinates which we collectively denote $x$. They must satisfy Cauchy-Riemann equations pair-wise: $\partial_i U^i = \partial_j U^j$  and $\partial_jU^i=-\partial_iU^j$, with $i\neq j$. No summation over repeated indexes is meant in this Section. It is straightforward to see, in an identical fashion as in the previous Section, that $\mathcal{O}\left(z^{\lfloor\frac{d+4}{2}\rfloor}\right)$ is the square-integrable part which coincides with the generators of proper diffeomorphisms and then everything goes along the same lines as before. 

However, in contrast with the $d=2$ case, for $d>2$ we can show that $\partial^3_{klm}U^j=0$ for all $ \,j,k,l,m$ (not necessarily different). In order to see this, first notice that Cauchy-Riemann equations imply that 
$ \partial_i^3 U^i = -\partial_j^3 U^j $. Then, grouping into threesomes $U^i, U^j, U^k$ with $i\neq j \neq k$, we get $ \partial_i^3 U^i = -\partial_j^3 U^j= \partial_k^3 U^k=-\partial_i^3 U^i=0 $. Hence, the $U^i$ functions are quadratic functions of the $x^i$ variable. Even more, by further using the Cauchy-Riemann equations it is possible to show that $U^i$ are at most quadratic in any variable: $\partial^2_j\partial_k U^k=\partial_j^2\partial_i U^i=-\partial^3_i U^i=0$. Also we have the following two identities: $\partial^2_k\partial_j U^k=-\partial^3_j U^k=\partial^2_j \partial_k U^j=-\partial_i^2 \partial_k U^j $ and $\partial^2_k\partial_j U^k=-\partial^2_i \partial_j U^k=\partial_i^2 \partial_k U^j$ which combined imply that $\partial^2_k\partial_j U^k=0$ and then also $\partial_j^3 U^k=0$. Vector fields generating improper diffs with functions $U^i$ at most quadratic in the boundary coordinates  are precisely the isometries of hyperbolic space. This completes the proof that the square integrability condition of metric perturbations imply that the asymptotic symmetry group is just the isometry group of hyperbolic space for $d>2$, which coincides with what is expected from the boundary conformal symmetries.

\section{Summary}

We have shown where are hidden in the one-loop partition function computation \cite{Giombi:One.Loop} both the imposition of Brown-Henneaux boundary conditions and the quotient by \textit{only} proper diffeomorphisms; the latter directly implying that improper diffs are symmetries which are exhibited in the final result as a character of the Virasoro group.

Let us describe now what we have found. First, the square-integrability of the metric perturbations demands that they fall off at infinity satisfying Brown-Henneaux asymptotic boundary conditions. Second, the  asymptotic vector fields that generate the allowed metric perturbations naturally split in two sets, 
\begin{itemize}
\item[] $\mathcal{V_P}:$ The set of square integrable vector fields. They have the same form as the generators of proper diffs in \cite{Brown.Henneaux:AdS3} but with a slightly milder fall off compatible with the more general analysis of \cite{Barnich.Brandt}. They generate gauge symmetries in the BRST formalism, and we checked that indeed they are adequate to implement the gauge condition on a metric perturbation.

\item[] $\mathcal{V_I}:$ The set of asymptotic vector fields modulo the set of square-integrable asymptotic vector fields. This is the vector space of generators of improper diffs in \cite{Brown.Henneaux:AdS3}. We checked that these vectors generate true symmetries in the sense that they keep perturbations in the chosen gauge, while they are not generated by ghost fields.   
\end{itemize}

Then, it is clear that by applying the BRST formalism together with a square-integrable heat kernel, the determinants involved in the one-loop partition function are taking into account both the Brown-Henneaux boundary conditions for metric perturbations and the quotient by only proper diffeomorphisms. Improper diffeomorphisms generate metric perturbations that remain contributing to the partition function. This is the reason why the Virasoro symmetry appears in the heat kernel computation. Actually, our presentation was motivated by the heat kernel method in \cite{Giombi:One.Loop}, but only relied on the fact of considering a space of square-integrable fields. Thus, the present manuscript permits to gain a deeper understanding of the computation of the gravity 1-loop partition function, such as in \cite{Martin:2019flv}, where determinants are obtained using a different method that assumes square-integrability.

In the last Section we extended the analysis to higher dimensions. We found that the square-integrability also accounts for the fact that in $D>3$ the asymptotic symmetry group is given by the isometry group in agreement with \cite{Henneaux:AdS4}.
  
We will dedicate a future work to extend the present approach to the case of pure gravity without cosmological constant \cite{Acosta:prep}.

\section*{Acknowledgments}
The authors thank Luca Ciambelli, Mart\'in Mereb and specially Gaston Giribet for enlightening discussions. ML thanks Edward Witten for correspondence. This work was partially supported by grants PIP and PICT from CONICET and ANPCyT.

\newpage
\bibliographystyle{toine}

\bibliography{3Dgravitybib}{}

\end{document}